\documentclass[12pt]{iopart}

\usepackage{graphicx}
\usepackage{dcolumn}
\usepackage{bm}

\begin{document}

\title[Magnetic properties of double exchange biased trilayers]{Magnetic properties of double exchange biased diluted magnetic alloy/ferromagnet/antiferromagnet trilayers}

\author{Carla Cirillo$^1$, Antoni Garc\'{i}a-Santiago$^{2,3}$, Joan Manel Hernandez$^{2,3}$, Carmine Attanasio$^1$ and Javier Tejada$^{2,3}$}

\address{$^1$ CNR-SPIN Salerno and Dipartimento di Fisica \lq\lq E.R. Caianiello\rq\rq, Universit\`{a} degli Studi di Salerno, Fisciano (Sa) I-84084, Italy\\
$^2$ Grup de Magnetisme, Departament de F\'{i}sica Fonamental, Facultat de F\'{i}sica, Universitat de Barcelona, c. Mart\'{i} i Franqu\`{e}s 1, planta 4, edifici nou, E-08028 Barcelona, Spain\\
$^3$ Institut de Nanoci\`{e}ncia i Nanotecnologia IN2UB, Universitat de Barcelona, c. Mart\'{i} i Franqu\`{e}s 1, planta 3, edifici nou, E-08028 Barcelona, Spain}

\date{\today}

\begin{abstract}
The magnetic properties of trilayers consisting of a diluted magnetic alloy, CuMn (Cu$_{0.99}$Mn$_{0.01}$), a soft ferromagnet, Py($\equiv$ Ni$_{0.8}$Fe$_{0.2}$), and an antiferromagnet, $\alpha$-Fe$_{2}$O$_{3}$, were investigated. The samples, grown by UHV magnetron sputtering, were magnetically characterized in the temperature range $T = 3-100$ K. Typical exchange bias features, namely clear hysteresis cycle shifts and coercivity enhancements, were observed. Moreover the presence of an inverse bias, which had been already reported for spin glass-based structures, was also obtained in a well defined range of temperatures and CuMn thicknesses.
\end{abstract}


\maketitle

\section{Introduction}

Exchange bias (EB) is a well known effect resulting from the exchange coupling at the interface between a ferromagnet (F) and an antiferromagnet (AF), once they are cooled through the ordering temperature of the latter. The key features of this mechanism are a shift and an enlargement of the magnetic hysteresis cycles of the system, compared to the ones corresponding to the single-F layer, which is in general accompanied by the so-called training effect. An extensive picture of EB is presented in several reviews devoted to this subject \cite{Schuller,Berkowitz,Stamps}. The EB effect is indeed observed in a large number of exchange coupled materials, such as conventional AF/F, ferrimagnet (ferri)/F \cite{Salazar}, and spin glass (SG)/F \cite{Marrows,Gruyters,Fiorani,Westerholt,Yuan1,Yuan2} systems, fabricated in a wide variety of forms, like nanoparticles, thin films, single crystals, and granular and disordered media. However, despite the large number of works devoted to the study of EB, the research in this field is far from being accomplished. Since its discovery \cite{MB1,MB2}, this phenomenon kept on gaining great attention due to its applications, mainly in magnetic storage and memory devices \cite{Fert,Grunberg}, but, also more recently, in the realization of superconducting spin switch valves \cite{Gu,Potenza,Birge,Leksin}. In this sense the fabrication of artificial thin films hybrids represents a unique tool to design systems with tunable coercive fields and pinning strength and to explore EB in several coupled materials. This research is also moved forward by the continuous developments in the fabrication techniques, which make possible the reliable realization of complex heterostructures consisting of layers of different materials only a few nanometers thick. However, the performance of these devices crucially depends on a large number of specific characteristics, such as interface properties (roughness, interdiffusion), crystallinity, and grain size \cite{Ali,Hussain,Kohn}.

In this paper the magnetic properties of double exchange biased systems consisting in DF/F/AF trilayers are reported. Here DF is a diluted magnetic CuMn alloy, with composition Cu$_{0.99}$Mn$_{0.01}$, F is permalloy (Py $\equiv$ Ni$_{0.8}$Fe$_{0.2}$), that is a soft ferromagnet, and AF is a native thermal oxide, namely hematite ($\equiv$ $\alpha$-Fe$_{2}$O$_{3}$), due to the natural oxidation of Py. This system enables one to explore the pinning exerted on the interleaved Py film by the external layers in which two different orderings are present. Several reasons make this kind of systems interesting. So far, the only work reporting on EB in Cu$_{0.94}$Mn$_{0.06}$/F bilayers \cite{Marrows} revealed unusual effects related to the SG nature of the CuMn layer, such as an inversion of the bias field in a well defined temperature region. The same result was also confirmed in a SG/F system where AuFe was the SG pinning layer \cite{Yuan1}. Here we investigate the problem working on the diluted side of the CuMn phase diagram where, as it will be seen in the following, the magnetic behavior is dominated by frustrated clusters and therefore the system cannot be considered a canonical SG. On the other hand, hematite is considered very attractive for EB applications because of both large giant magnetoresistance ratios and high bulk N\'{e}el temperature \cite{Bae1,Blamire}. In particular, being the hematite layer a native oxide film here, the resulting trilayer comes effectively from the fabrication of only two layers. Finally, the proposed trilayers can provide a controllable twisted spin structure \cite{Guo}, which could be interesting for the realization of spin-based devices.

This work is organized as follows. After a brief description of the samples fabrication, the results of their characterization both by high- and low-angle x-ray measurements are presented. A preliminary detailed magnetic characterization was performed on single CuMn films, as well as on protected and unprotected Py films. The central part of the paper is devoted to the investigation of the exchange bias and coercivity of CuMn/Py/$\alpha$-Fe$_{2}$O$_{3}$ trilayers as a function of the CuMn thickness. From these measurements the values of the interfacial exchange energies for the two pinning layers, namely CuMn and $\alpha$-Fe$_{2}$O$_{3}$, are evaluated. A negative exchange field is reported in an intermediate CuMn thickness range and in a well defined temperature window.

\section{Experiment}

\subsection{Sample fabrication}

CuMn/Py bilayers were deposited on Si(100) substrates by UHV dc diode magnetron sputtering at room temperature, at a base pressure in the low $10^{-8}$ mbar range and at processing Ar pressure of 3$\times$$10^{-3}$ mbar. The typical deposition rates were 0.18 nm/s for CuMn and 0.24 nm/s for Py, as measured by a quartz crystal monitor previously calibrated by low-angle x-ray reflectivity measurements (XRR) on deliberately deposited thin films of each material. Samples have constant Py thickness ($d_{\mathrm{Py}}$ = 12 nm) and variable CuMn ones ($d_{\mathrm{CuMn}} = 4 - 57$ nm). In order to obtain samples under identical deposition conditions four different substrates were loaded in the deposition chamber. A movable protecting shutter driven by a computer controlled step motor allows to fabricate hybrids having one layer with constant thickness, changing the thickness of the other as desired. After the deposition of the CuMn/Py bilayers, the samples were exposed to atmospheric pressure at room temperature. As it as been unambiguously demonstrated, the thermal oxidation of the external Py film exposed to the laboratory environment produces an
$\alpha$-Fe$_{2}$O$_{3}$ surface AF layer \cite{Bailey,Bajorek,Fulcomer,Pollak,Flokstra}. Indeed, according to the Fe-O phase diagram \cite{Muan}, $\alpha$-Fe$_{2}$O$_{3}$ is the stable phase at room temperature and atmospheric pressure. Therefore we can assert that the final structure of the investigated sample is CuMn/Py/$\alpha$-Fe$_{2}$O$_{3}$.

Single Py and CuMn films of various thicknesses were also fabricated for reference. In order to perform a comparative study to discern the role played by the hematite layer, single Py films with a protective Nb cap layer, 2-nm-thick, were also prepared. All depositions were performed in zero magnetic field. Energy dispersion spectroscopy (EDS) analysis confirmed the nominal Mn concentration of the CuMn sputtering target source, namely 0.01, in the deposited films.

\subsection{Structural characterization}

\begin{figure}[!ht]
\centering
\includegraphics[scale=0.8]{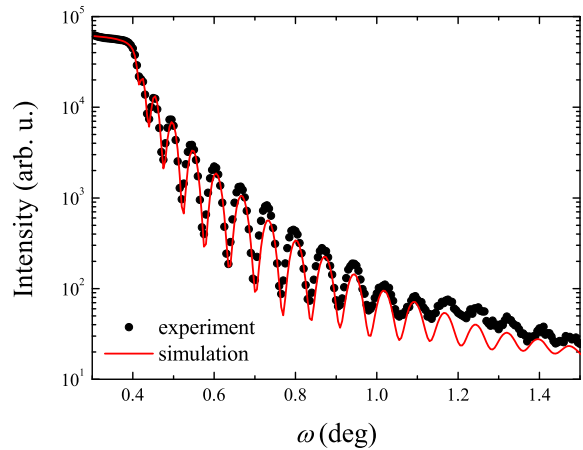}
\caption{(Colour online). Experimental (scatter) and simulated (line) specular XRR profile of a CuMn/Py/$\alpha$-Fe$_{2}$O$_{3}$ trilayer.}
\label{XRR}
\end{figure}

In order to investigate the interface quality and the layering of the structures, preliminary XRR measurements were performed on selected samples using a Philips X'Pert-MRD high resolution diffractometer equipped with a four-circle cradle and a CuK$_\alpha$ source. Moreover, it is well established that XRR provides high sensitivity to thin surface layers and therefore it can return information concerning the composition of the surface oxide layer. Technical details concerning the measurements can be found elsewhere \cite{XRR_KL}. In Fig. \ref{XRR} the specular XRR profile of a CuMn/Py/$\alpha$-Fe$_{2}$O$_{3}$ trilayer is shown. The data (circles) are treated quantitatively by computer-aided simulation (line) \cite{Parratt,Nevot} in order to extract values for interface roughness, layer thickness, and external layer composition. In particular, the experimental profile is not compatible with the presence of a surface Ni oxide, while it is satisfactorily reproduced by a theoretical model which uses the layer thicknesses and the corresponding root-mean-square roughnesses as fitting parameters. Such model allowed to obtain $d_{\mathrm{CuMn}}$ = 38.7 nm, $d_{\mathrm{Py}}$ = 14.7 nm, and $d_{\mathrm{\alpha-Fe_{2}O_{3}}}$ = 1.2 nm for the thicknesses, and $\sigma_{\mathrm{CuMn}}$ = 0.6 nm, $\sigma_{\mathrm{Py}}$ = 1.5 nm, and $\sigma_{\mathrm{\alpha-Fe_{2}O_{3}}}$ = 0.5 nm for the roughnesses. This result confirms the presence of the most stable Fe oxide, $\alpha$-Fe$_{2}$O$_{3}$, on top of the Py layer, and returns values for the hematite thickness which are consistent with compositional depth profiles determined by x-ray photoemission spectroscopy (XPS) on unprotected Py films \cite{Pollak}. Moreover XRR reveals smooth interfaces, which promote an efficient exchange coupling between the different layers \cite{Bae1,Bae2,Bae3}.

High-angle $\theta$-2$\theta$ measurements were performed using a standard x-ray powder diffractometer with CuK$_\alpha$ radiation in order to study the crystallographic orientation of the different layers. The diffraction patterns acquired for the CuMn/Py/$\alpha$-Fe$_{2}$O$_{3}$ trilayers (not reported here) indicate that both Cu and Py present an fcc oriented growth in the (111) direction and show, in the limit of our experimental accuracy, that no other crystallographic phases are present in the samples. Finally, the experimental data do not show any peak
related to the presence of either the thin $\alpha$-Fe$_{2}$O$_{3}$ layer or the extremely diluted Mn atoms.

\section{Results and discussion}

\subsection{Single CuMn and Py films}

All the magnetic measurements reported in this subsection and in the following were performed in a commercial superconducting quantum interference device magnetometer \cite{MPMS} with the magnetic field always applied parallel to the substrate plane. To address the problem of the sample positioning with respect to the magnetometer pick-up coils, prior to all the data acquisition a centering procedure (at $T$ = 5 K and in a magnetic field that never exceeded $H =$ 500 Oe) was performed. The temperature was then increased to $T$ = 100 K and magnetic hysteresis cycles, $m(H)$, were measured at different temperatures cooling the samples in a field of -500 Oe down to each target temperature. The samples were always heated up to $T$ = 100 K between two subsequent magnetic cycles.

A magnetic characterization of single CuMn films was preliminary performed. CuMn was chosen among others because it is a well studied system whose properties had been investigated in a broad range of compositions \cite{Mydosh,Gibbs}, spanning from diluted \cite{Ruitenbeek,deJongh} to concentrated regimes \cite{Kouvel,Ziq,Webb} with a particular focus on the SG ordering \cite{Kenning,Hoines,Kenning2,Chu}. As in all the SG alloys, the randomness in the position of the magnetic impurities and in the sign of the neighboring couplings in CuMn generate frustration and random distribution of coupling strengths and directions below a freezing temperature, $T_{f}$, and results in the well-known Ruderman-Kittel-Kasuja-Yosida (RKKY) interaction \cite{Mydosh}. Among many peculiar properties, concentrated SG alloys also exhibit EB intrinsically \cite{Mydosh,Kouvel,Ziq,Webb}, an effect which was first explained by modelling the alloys as systems of mutually interacting F and AF assemblages arising from inhomogeneous impurity distribution \cite{Kouvel}. Indeed further studies revealed the formation of chemical clusters of Mn atoms in CuMn alloys \cite{deJongh}, which can be thus described as an ensemble of fairly large F clusters interacting with each other in a
randomized SG way \cite{deJongh,Webb}. These effects are important even for Mn concentrations of the order of 1 percent, so that diluted alloys may still be considered frustrated systems \cite{deJongh}.

\begin{figure}[!ht]
\centering
\includegraphics[scale=0.8]{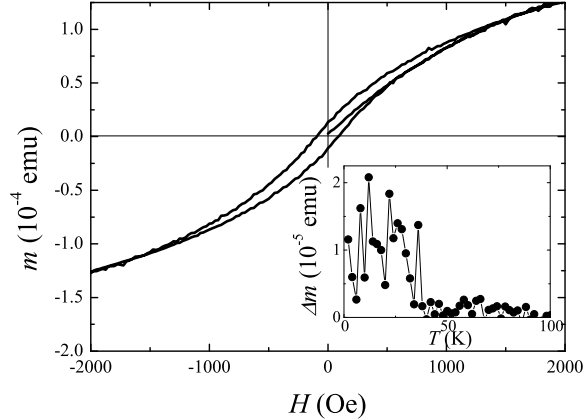}
\caption{Magnetic hysteresis cycle for a 230-nm-thick CuMn film cooled down to $T$ = 3 K in zero magnetic field. Inset: temperature dependence of
the difference between the field cooled and the zero-field cooled magnetic moment, $\Delta m = m_{\rm{FC}} - m_{\rm{ZFC}}$.} \label{MTMHCM}
\end{figure}

In Fig. \ref{MTMHCM} the zero-field cooled $m(H)$ cycle for a single 230-nm-thick CuMn film at $T$ = 3 K is reported. The magnetization value at the maximum applied field is $M_{\mathrm{CuMn}}^{\mathrm{max}}$ $\approx$ 10 emu/cm$^{3}$. Both the reduced coercivity and the lack of saturation even at fields up to $H$= 2$\times$10$^{4}$ Oe (not shown here), are reminiscent of a SG behavior \cite{Mydosh}. However, when measuring the zero-field cooled ($m_{\rm{ZFC}}$) and the field cooled ($m_{\rm{FC}}$) magnetic moment as a function of temperature to determine the value of $T_f$, instead of observing a SG peak an onset of magnetic irreversibility is detected at $T_{\rm{irr}}$ $\approx$ 40 K. This is shown in the inset of Fig. \ref{MTMHCM}, where the difference $\Delta m \equiv m_{\rm{FC}}- m_{\rm{ZFC}}$ is plotted versus $T$ for a cooling field of $H =$ 500 Oe. The small magnetic signal is responsible for the scattering of the data points. Nevertheless, taking into account the low Mn concentration measured by EDS analysis, this $T_{\rm{irr}}$ value is unexpectedly high compared to the values reported for $T_f$ for CuMn thin films \cite{Kenning,Hoines}. This result can be indeed interpreted considering $T_{\rm{irr}}$ as the blocking temperature of Mn clusters below which all of them are freezed in random directions. The goal is to understand whether this configuration can give rise to the EB effect just like a canonical SG system does \cite{Marrows}.

\begin{figure}[!ht]
\centering
\includegraphics[scale=0.8]{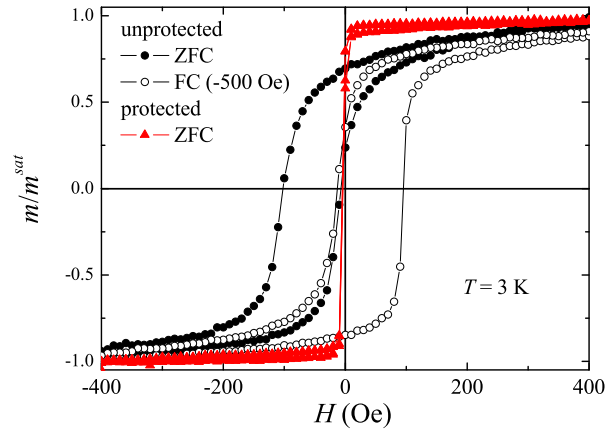}
\caption{(Colour online). Magnetic hysteresis cycles at $T$ = 3 K for an unprotected Py film, namely a Py(12)/$\alpha$-Fe$_{2}$O$_{3}$(1.2) bilayer, measured with a ZFC (solid dots) and a negative FC (open dots)
procedure, and for a ZFC protected Py film, namely a Py(12)/Nb(2) bilayer (solid triangles).}
\label{SP-SPN}
\end{figure}

Magnetic hysteresis cycles were also measured on both unprotected and protected Py single films at $T$ = 3 K as reported in Fig. \ref{SP-SPN}. The normalized hysteresis loop measured on the protected sample, namely the Py(12)/Nb(2) bilayer (the numbers in brackets indicate the layer thickness in nm), presents almost no coercivity and, moreover, is centered at the origin. These results are expected for a soft F material free to reverse in an external applied field. As discussed in the previous section, the unprotected sample presents an external hematite layer. Hematite is a weak F system at temperature higher than the so-called Morin temperature, $T_{M}$, while it behaves like a uniaxial AF underneath \cite{delBarco}. In hematite particles $T_{M}$ strongly depends on their size \cite{Fiorani_hematite}, hence a thickness dependence of $T_{M}$ is also expected in thin films. From Fig. \ref{SP-SPN} it follows that the Py(12)/$\alpha$-Fe$_{2}$O$_{3}$(1.2) bilayer exhibits features typical of the EB effect, that is a broadened $m(H)$ curve shifted from the origin. This is shown in Fig. \ref{SP-SPN} where the open (closed) circles are the points acquired after a FC with $H$= -500 Oe (ZFC) process. The two loops are symmetrical, the sign of the bias
field being opposite in the two cases. The shifted hysteresis cycle resulting from the ZFC procedure is not surprising since, due to the measurement protocol discussed at the beginning of this subsection, the cooling procedure started from a positive remanent state of the F layer \cite{OGrady} and indicates that the field of Py at the remanence is strong enough to induce exchange anisotropy. Indeed the state of the magnetization of the F layer strongly affects the exchange bias in F/AF systems \cite{Miltenyi}.

\begin{figure}[!ht]
\centering
\includegraphics[scale=0.8]{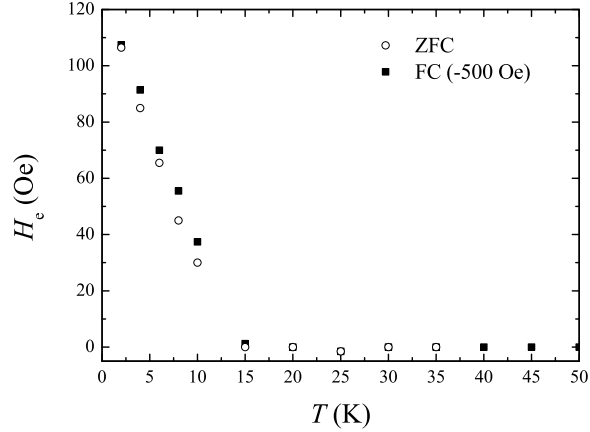}
\caption{Exchange bias field, $H_{e}$, as a function of the temperature, cooling the sample in zero field (open symbols) and in the presence of a field $H$ = -500 Oe (closed symbols).}
\label{Tt}
\end{figure}

From the values of the coercive fields of the cycles, $H_{c1}$ and $H_{c2}$, the values of both the exchange bias field, $H_{e}$= ($H_{c1}$+$H_{c2}$)/2, and sample coercivity, $H_{c}$= ($H_{c1}$-$H_{c2}$)/2, were calculated (note that the numbers quoted in the following for $H_{c1}$ and $H_{c2}$ are not absolute values). For the Py/$\alpha$-Fe$_{2}$O$_{3}$ bilayer in Fig. \ref{SP-SPN} it is $H_{e}$ = 55 Oe and $H_{c}$ = 47 Oe. To evaluate quantitatively the exchange coupling strength exerted from the $\alpha$-Fe$_{2}$O$_{3}$ layer on the Py one, the interface exchange energy per unit area can be defined as $\Delta E$ = $M_{\mathrm{F}}$$d_{\mathrm{F}}$$H_{e}$, where $M_{\mathrm{F}}$ and $d_{\mathrm{F}}$ are respectively the saturation magnetization and the thickness of the F layer \cite{Schuller}. In this case it is $\Delta E_{\mathrm{\alpha-Fe_{2}O_{3}}}$ = 0.025 erg/cm$^{2}$. This value, as well as the ones for $H_{c}$ and $H_{e}$, is consistent with the numbers reported in literature for this system \cite{Bae2,Bae3}. The temperature dependence of the exchange bias field for the Py/$\alpha$-Fe$_{2}$O$_{3}$ bilayer, following both the ZFC and the FC procedures, is studied in Fig. \ref{Tt}. From these measurements it can be inferred that the thin $\alpha$-Fe$_{2}$O$_{3}$ layer produces a bias field when cooled below a transition temperature $T_{t}$ $\approx$ 15 K in accordance with Refs. \cite{Bailey,OGrady}, since above this temperature both the loop shift and the enhanced coercivity (not shown here) disappear. Moreover the data confirm a weak dependence of the bias field on the cooling process.

\begin{figure}[!ht]
\centering
\includegraphics[scale=0.8]{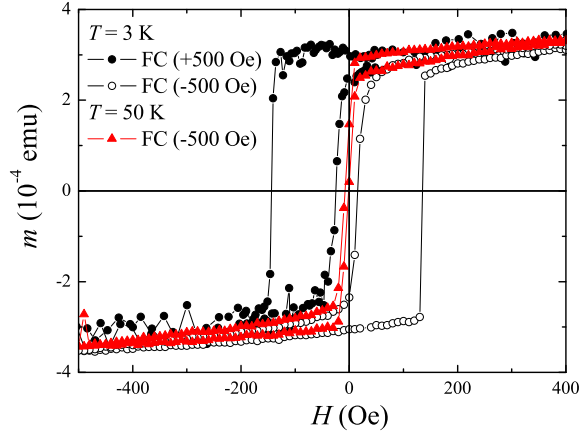}
\caption{(Colour online). Magnetic hysteresis cycles measured on the CuMn(56)/Py(12)/$\alpha$-Fe$_{2}$O$_{3}(1.2)$ trilayer after a FC procedure at $T$ = 3 K cooling the sample in a field $H$ = $\pm$500 Oe (solid and open dots for plus and minus signs, respectively), and at $T$ = 50 K cooling the sample in a field $H$ = -500 Oe (solid triangles).}
\label{PCM8}
\end{figure}

It is worth mentioning that the exchange bias strongly depends on the spin structure at the interface, especially on the angle between the F and AF spins. However in this paper the interfacial spin configuration of the hematite layer \cite{Blamire} was not investigated while, due to the reduced Py thickness, an in-plane easy axis for the Py layer was assumed instead \cite{Trunk}.

\subsection{CuMn/Py/$\alpha$-Fe$_{2}$O$_{3}$ trilayers}

\begin{figure}[!ht]
\centering
\includegraphics[scale=0.8]{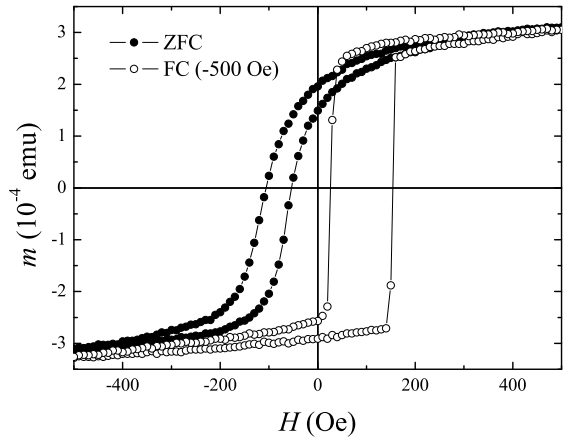}
\caption{Solid (open) dots represent the magnetic hysteresis cycle measured on the CuMn(15.5)/Py(12)/$\alpha$-Fe$_{2}$O$_{3}(1.2)$ trilayer after a ZFC (FC) procedure at $T$ = 3 K.}
\label{PCM2}
\end{figure}

In Fig. \ref{PCM8} the $m(H)$ curves acquired at $T$ = 3 K for the CuMn(56)/Py(12)/$\alpha$-Fe$_{2}$O$_{3}(1.2)$ trilayer are reported. The sample was cooled from 100 K down to 3 K in a magnetic field of either $H$= 500 Oe (solid dots) or $H$ = -500 Oe (open dots). Both loops are shifted with respect to the origin, namely for a FC procedure in a positive (negative) magnetic fields the loop shifts towards negative (positive) fields, the $m(H)$ curves being perfectly symmetrical. In particular it is $H_{e}$ = 79 Oe and $H_{c}$ = 59 Oe. The triangles indicate the cycle measured at $T$ = 50 K, that is at $T > T_{\mathrm{irr}}$, after a FC at $H$ = -500 Oe. As it can be inferred in this case, both the enhanced coercivity and the loop shift disappear. The effects of the two different pinning layers are evidenced in Fig. \ref{PCM2}, where the hysteresis loops at $T$ = 3 K after ZFC and FC procedures in a field of $H$ = -500 Oe are reported for the trilayer with $d_{\mathrm{CuMn}}$ = 15.5 nm. The ZFC cycle is elongated and it is centered at $H_{e}$ = -79 Oe. Recalling the result presented in Fig. \ref{SP-SPN}, namely the shift of the loop towards negative field after a ZFC procedure from a remanent state of the ferromagnet, it follows that the $\alpha$-Fe$_{2}$O$_{3}$ layer is also responsible for the ZFC shift of Fig. \ref{PCM2}, with a corresponding interfacial exchange bias energy $\Delta E_{\mathrm{\alpha-Fe_{2}O_{3}}}$ = 0.048 erg/cm$^{2}$. On the other hand, after the FC the loop is much sharper and the pinning direction is reversed, being $H_{e}$ = 90 Oe. Also in this case it is possible to estimate the interfacial energy strength, which now is the sum of the energies exerted at the two interfaces CuMn/Py and Py/$\alpha$-Fe$_{2}$O$_{3}$, namely $\Delta E_{\mathrm{CuMn,\alpha-Fe_{2}O_{3}}}$ = $\Delta E_{\mathrm{\alpha-Fe_{2}O_{3}}}$+$\Delta E_{\mathrm{CuMn}}$ \cite{Guo,Sort}. It results that $\Delta E_{\mathrm{CuMn,\alpha-Fe_{2}O_{3}}}$ = 0.054 erg/cm$^{2}$ and therefore $\Delta E_{\mathrm{CuMn}}$ $\approx$ 0.006 erg/cm$^{2}$. The same evaluation performed on the sample with $d_{\mathrm{CuMn}}$ = 56 nm gives $\Delta
E_{\mathrm{\alpha-Fe_{2}O_{3}}}$ = 0.041 erg/cm$^{2}$ and $\Delta E_{\mathrm{CuMn,\alpha-Fe_{2}O_{3}}}$ = 0.053 erg/cm$^{2}$, consequently $\Delta E_{\mathrm{CuMn}}$ $\approx$ 0.012 erg/cm$^{2}$. This last result indicates that in this $d_{\mathrm{CuMn}}$ range the contribution of the CuMn layer increases with its thickness. Moreover, the values estimated for $\Delta E_{\mathrm{CuMn}}$ confirm that the energy scales are much smaller in magnitude in SG compared to AF \cite{Marrows}. It is worth underlining that in this kind of trilayers it is also possible to change the relative
orientation of the coupling \cite{Guo,Sort} between Py and the two pinning layers in order to realize a defined spin structure in the F layer. This issue has not been investigated in this work and will be the subject of future studies.

\begin{figure}[!ht]
\centering
\includegraphics[scale=0.8]{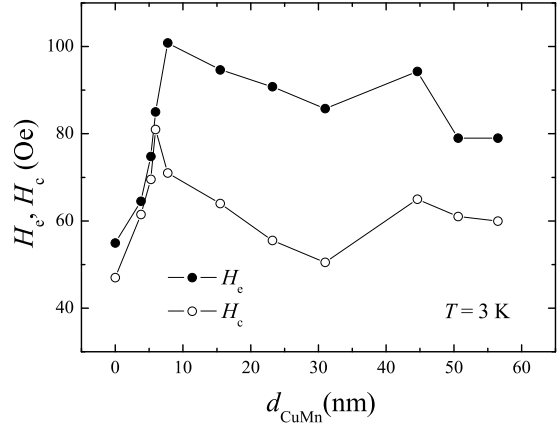}
\caption{Exchange bias field, $H_{e}$ (solid dots), and coercivity, $H_{c}$ (open dots), of CuMn/Py/$\alpha$-Fe$_{2}$O$_{3}$ trilayers as a function of the CuMn thickness, $d_{\mathrm{CuMn}}$, at $T$ = 3 K.}
\label{HeHc}
\end{figure}

\begin{figure}[!ht]
\centering
\includegraphics[scale=0.8]{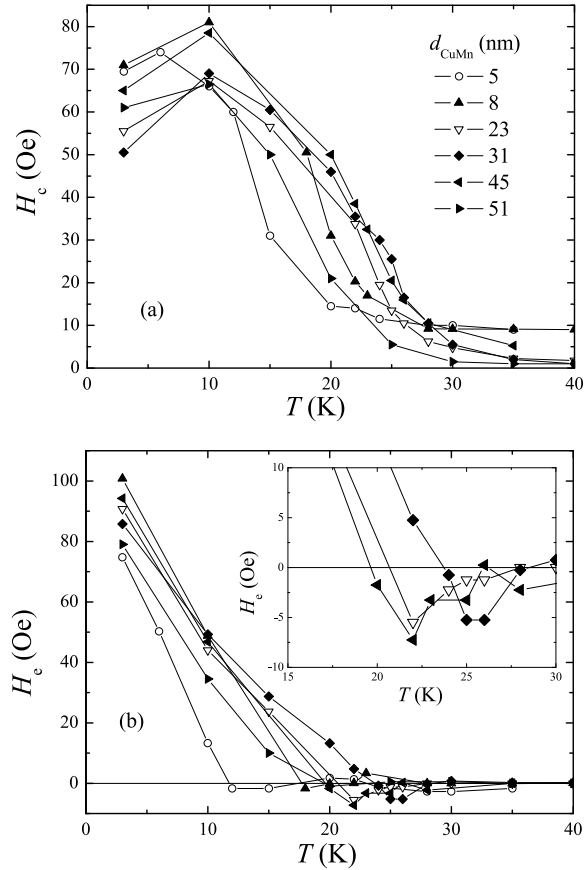}
\caption{Temperature dependence of (a) the coercivity, $H_{c}$, and (b) the exchange bias field, $H_{e}$, for a representative of CuMn/Py/$\alpha$-Fe$_{2}$O$_{3}$ trilayers. The inset of panel (b) shows an enlargement of the $H_{e}(T)$ behavior for the trilayers with $d_{\mathrm{CuMn}}$ = 23 nm (downward triangles), 31 nm (diamonds), and 45 nm (left-pointing triangles).}
\label{HeHcT}
\end{figure}

The hysteresis loops were systematically acquired for all the trilayers with variable CuMn thickness. The dependence of both the bias field and the coercivity on $d_{\mathrm{CuMn}}$ at $T$ = 3 K after a FC process in a field of $H$ = -500 Oe from $T$ = 100 K is shown in Fig. \ref{HeHc}. The curves do not show a strong thickness dependence, except for $d_{\mathrm{CuMn}}$ $\leq$ 6 nm, where both $H_{e}$ and $H_{c}$ decrease as $d_{\mathrm{CuMn}}$ diminishes. Therefore, as reported for the CuMn/Co system, the minimum CuMn thickness for EB to occur lies in this region \cite{Marrows}. The $H_{c}(d_{\mathrm{CuMn}})$ behavior also seems to exhibit the same non-monotonic trend as in CuMn/Co bilayers \cite{Marrows}, while a monotonic dependence was reported for the AuFe/Py system \cite{Yuan1}. The dependences observed in Fig. \ref{HeHc} are in qualitative agreement with those recently obtained in a theoretical work devoted to EB features in SG/F bilayers \cite{Usadel1}.

\begin{figure}[!ht]
\centering
\includegraphics[scale=0.8]{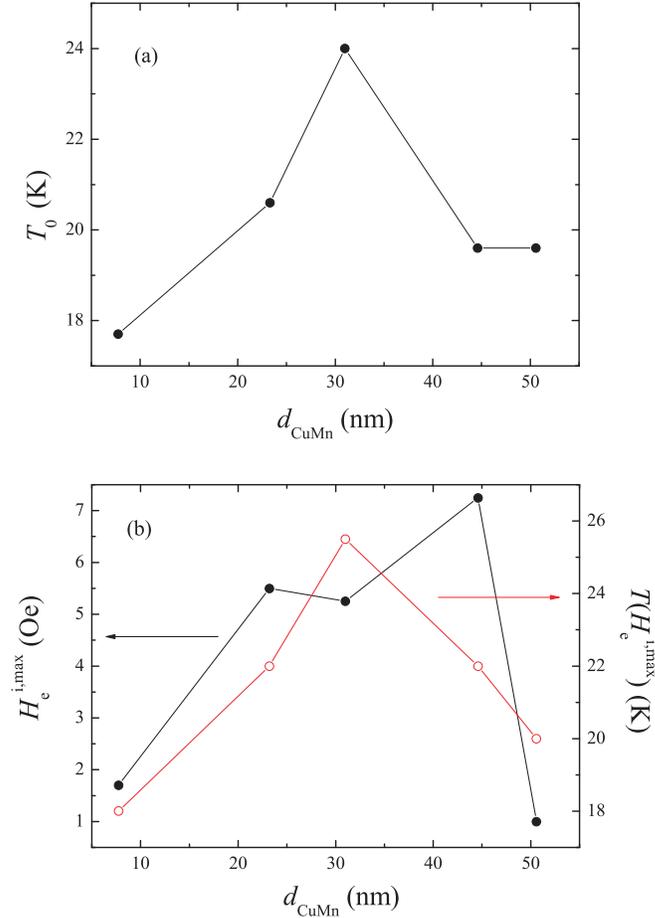}
\caption{(Colour online). CuMn thickness dependence of (a) the blocking temperature, $T_{0}$, and (b) the maximum inverse bias field, $H_{e}^{i,\rm{max}}$ (solid dots), and the temperature at which the maximum effect is present, $T(H_{e}^{i,\rm{max}})$ (open dots), for a representative of CuMn/Py/$\alpha$-Fe$_{2}$O$_{3}$ trilayers.}
\label{inverse}
\end{figure}

The temperature dependence of $H_{c}$ and $H_{e}$ for a representative of CuMn/Py/$\alpha$-Fe$_{2}$O$_{3}$ trilayers with different $d_{\mathrm{CuMn}}$ values is shown in Fig. \ref{HeHcT}. The $H_{c}(T)$ behavior (Fig. \ref{HeHcT}a) shows a maximum at $T$ $\approx$ 10 K, in qualitative agreement with the one reported for the CuMn/Co system \cite{Marrows}. In the present study, however, the peak is almost thickness independent, and seems not to be related to the corresponding $H_{e}(T)$ curves, as claimed in Ref. \cite{Marrows}. The curves saturate at a temperature compatible with $T_{\rm{irr}}$, a feature common to other SG/F systems \cite{Marrows,Yuan1,Yuan2}. More interesting is the temperature dependence of the bias field, which decreases as $T$ increases, independently of the CuMn thickness. However, in a well defined $d_{\mathrm{CuMn}}$ range (23-45 nm), a sign change in the loop shift is present just below the blocking temperature $T_{0}$, namely the temperature at which EB disappears. This effect is highlighted in the inset of Fig. \ref{HeHcT}b for the three samples with $d_{\mathrm{CuMn}}$ = 23, 31, and 45 nm. The moderate cooling fields applied in these experiments rule out the possibility that this inverse EB is of the nature of the one reported for AF/F interfaces \cite{Schuller}. In contrast, this effect seems to be a direct consequence of the nature of the pinning layer, as recently observed also in other SG/F systems \cite{Marrows,Westerholt,Yuan1}. This conclusion is also supported by the result presented in Fig. \ref{Tt}, which indicates that the pinning effect exerted from the hematite layer vanishes around 15 K.

According to the mean field model suggested in Ref. \cite{Marrows}, the RKKY interaction is the key ingredient to observe the negative EB in SG/F systems. Indeed, increasing the Mn concentration, and hence the strength of nearest neighbors interactions compared to the RKKY one, it is possible to tune the amplitude of the effect to suppress it even completely. As a matter of fact, the aforementioned model can account for both the temperature and the thickness dependences of the EB field. Here the study performed in Ref. \cite{Marrows} has been extended to the regime of diluted alloys, demonstrating that EB and its inversion are produced also at very low Mn concentrations. The main features of the negative effect observed in this work are summarized in Fig. \ref{inverse}. In panel (a) a non-monotonic thickness dependence of the blocking temperature is observed with a maximum of $T_{0}^{\rm{max}}$ = 24 K at $d_{\mathrm{CuMn}}$ = 31 nm, in contrast with the results observed for both CuMn/Co and AuFe/Py bilayers, where $T_{0}$ is a monotonously increasing function of $d_{\mathrm{CuMn}}$ \cite{Marrows,Yuan1}. In panel (b) the thickness dependence of the maximum inverse field, $H_{e}^{i,\rm{max}}$ (solid dots), and the temperature at which the maximum effect is present, $T(H_{e}^{i,\rm{max}})$ (open dots), are reported. Both curves act also in a non-monotonic way, their behaviors mimicking that of $T_{0}(d_{\mathrm{CuMn}})$. The measured values of the negative EB field are consistent with those reported in Ref. \cite{Marrows}.

\section{Conclusions}

The exchange bias effect in CuMn/Py/$\alpha$-Fe$_{2}$O$_{3}$ trilayers was explored in this work. The results include the CuMn thickness and the temperature dependences of both the exchange bias and the coercive fields of the magnetic hysteresis cycles. The effective interfacial exchange coupling energy resulting from the two interfaces, namely CuMn/Py and Py/$\alpha$-Fe$_{2}$O$_{3}$, was also estimated. A negative exchange bias effect was observed in a CuMn thickness region between 23 and 45 nm, just below the blocking temperature $T_{0}$. In this system the interest in fundamental aspects of exchange coupling is closely linked with the requirement of tunable and versatile magnetization configurations in hybrid thin film structures. The use of a diluted magnetic alloy gives the opportunity to study the effect of random clustering with a distribution of sizes, shapes and anisotropies on the exchange bias phenomenon. The system presented here is therefore more complex compared to the ones that employ canonical spin glasses containing isolated, randomly oriented moments dispersed in a metallic matrix, which were reported in previous works \cite{Marrows,Yuan1,Yuan2}. However, despite this complexity the peculiar features that are typical of spin glass-based structures, such as the observed negative bias field, were also observed. Hence this work confirms the crucial role played by disorder and frustration in the exchange bias effect, as it was also reported for diluted antiferromagnetic systems \cite{dilutedAF}.

In addition to the contribution to the comprehension of the microscopic origin of exchange bias, the proposed trilayer structures are also interesting because, combining the pinning at two different interfaces, are suitable for realizing a controlled spin structure in the ferromagnetic layer. In particular, taking advantage of the different magnetic orderings of the pinning layers it can be possible to tune the magnetization configuration as a function of both temperature and magnetic field with possibly important consequences also on the electronic transport \cite{Guo}. In this sense the advantage of the proposed system is to exploit the desired native oxide of Py as an external source for
pinning. This implies both high quality interface with the adjacent ferromagnetic layer and simplification of the fabrication procedures.

\ack

C. C. acknowledges the CNR for the financial support within the CNR Short Term Mobility Program. A. Vecchione and R. Fittipaldi are gratefully acknowledged for low angle XRR and EDS analysis, respectively. C. C. and A. G.-S. wish to thank D. Fiorani for fruitful discussion and careful reading of the manuscript. A. G.-S. and J. M. H. thank Universitat de Barcelona for backing their research. J. T. appreciates financial support from ICREA Academia. This work was funded by the Spanish Government projects MAT2008-04535 and MAT2011-23698.

\end{document}